\documentstyle[prl,aps,twocolumn,epsf]{revtex}
\draft
\epsfclipon

\begin{document}

\twocolumn[\hsize\textwidth\columnwidth\hsize\csname
@twocolumnfalse\endcsname
\title{Proximity effects and Andreev reflection in mesoscopic
SNS junction with perfect NS interfaces}

\author{Z.~D.~Kvon,$^1$ T.~I.~Baturina,$^1$ R.~A.~Donaton,$^2$ 
M.~R.~Baklanov,$^{1,2}$ K.~Maex,$^2$ E.~B.~Olshanetsky,$^1$ 
A.~E.~Plotnikov,$^1$ J.~C.~Portal$^3$}

\address{$^1$ Institute of Semiconductor Physics, 630090, Novosibirsk, Russia}
\address{$^2$ IMEC, Kapeldreef 75, B-3001 Leuven, Belgium}
\address{$^3$ CNRS-LCMI, F-38042, Grenoble, France}

\maketitle

\begin{abstract}
Low temperature transport measurements on superconducting film--normal
wire--superconducting film (SNS) junctions fabricated on the basis of
thin superconducting polycrystalline PtSi films are reported. Due to
the perfectness of SN boundaries in the junctions, zero bias 
resistance dip
related to pair current proximity effect and subharmonic energy gap
structure originating from phase coherent multiple Andreev reflections have
been observed  and studied.
\end{abstract}

]
\bigskip
\narrowtext

Research activity in the study of properties of NS and SNS
junctions (where N is a normal metal and S is a superconductor) has
increased significantly over the past few years, mainly owing to
technological advances in fabrication of mesoscopic hybrid systems.
Until now, all such systems have been made by a 
combination of different materials, for example, the 
superconductor--normal metal pairs Al-Ag~\cite{1}, Nb-Al and Nb-Ag~\cite{2}, 
superconductor--heavily doped semiconductor pairs Nb-$n^+$InGaAs~\cite{3}, 
Nb-$n^+$InAs~\cite{4},
Nb-$p^+$Si~\cite{5,6}, Al-$n^+$GaAs~\cite{7} and so on. 
One of the important topics of
the physics of NS and SNS junctions is the current--voltage
characteristics behavior. Two interesting phenomena can be observed
from the dependence of 
SNS differential resistance versus bias
voltage ($dV/dI$-$V$): an anomalous resistance dip at zero 
bias~\cite{2,3,4,5,6,7}, the
so-called zero bias anomaly (ZBA); and symmetrical dips at nonzero
biases~\cite{7,8}. 
In some cases the existence of tunnel barriers at the NS interfaces
does not give the
possibility of the definite interpretation of experimental results.

In this paper we report the fabrication and study of SNS junctions
with perfect SN interfaces. The best way to obtain a perfect junction
is to have both superconducting and normal metal parts of an SNS
junction made of the same material. It is known that the wires made from
some thin superconducting metallic films can have the transition
temperature $T_c$ less than that of the film itself. We have used this
property to fabricate the ideal SNS junctions.

We started with fabrication of ultrathin PtSi films. Smooth,
continuous and uniform PtSi films with thickness of 6~nm were formed
on Si substrates by deposition of a thin Pt layer followed by a rapid
thermal annealing step at 450$^\circ$C for 60~s to convert the as-deposited
Pt into PtSi. The films formed were polycrystalline, with an average
grain size of roughly 20~nm.

The samples used in the experiments are three Hall bridges
with $50~\mu$m in width and $100~\mu$m in length. 
The main parameters of the
films obtained from Hall measurements at $T>T_c=0.54$~K are presented in
Table~\ref{table1}. The diffusion constant is estimated assuming the
simple
free electron model. As is seen from Table~\ref{table1}, our PtSi
films are metal films with small mean free path. It should be noted that
they have a hole type conductivity.

\begin{figure}
\centerline{\epsfxsize2.6in\epsfbox{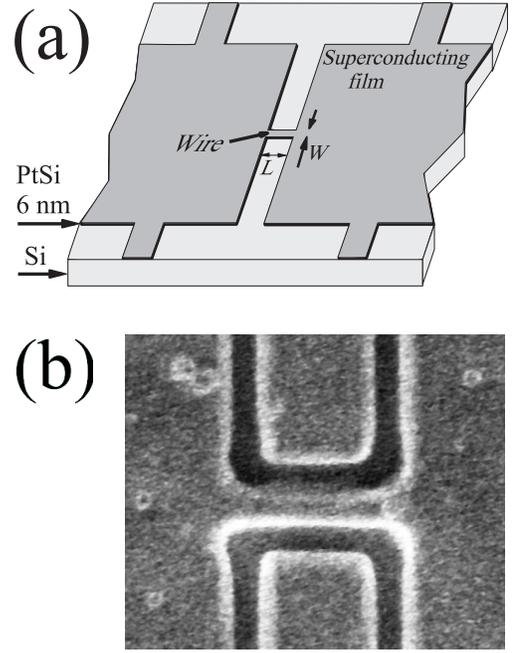}\bigskip}
\caption{(a) Schematic view of a junction. (b) SEM image of the PtSi
wire with length $1.5~{\mu}$m and width $0.3~{\mu}$m formed by electron beam
lithography and subsequent plasma etching.}
\label{fig1}
\end{figure}

PtSi wires of length $L=1.5~{\mu}$m and $6~{\mu}$m and width $W=0.3~{\mu}$m 
were fabricated by means of electron lithography and subsequent plasma
etching and placed in one of the Hall bridges. 
For a schematic picture of the samples, and a 
scanning electron
micrograph of one of the structures, see Fig.~\ref{fig1}. At 
$T>T_c$ the
resistances of the wires 
are $R_N\sim610~\Omega$ for $L=1.5~\mu$m and
$R_N\sim2600~\Omega$ for $L=6~\mu$m. 
These values correspond to the 
wire sheet resistance of
$R_{\Box}=120$--$130~\Omega$, 
that is slightly larger than $R_{\Box}$
of the film itself. About ten samples were investigated. 
None of the wires were
superconducting down to $T=35$~mK. The reason for that is not
completely understood. The suppression of the superconductivity
possibly results from the intrinsic stresses in the PtSi films which
increase in constrictions. This suggestion is supported by the
enhancement of the PtSi sheet resistance. 
Maybe there is another reason of the losing superconductivity in 
the wires.
Nevertheless, after the processing we
have at $T<T_c$ the SNS junctions which consist of two
superconducting seas connected by the normal metal PtSi wire.

\begin{figure}
\centerline{\epsfxsize2.6in\epsfbox{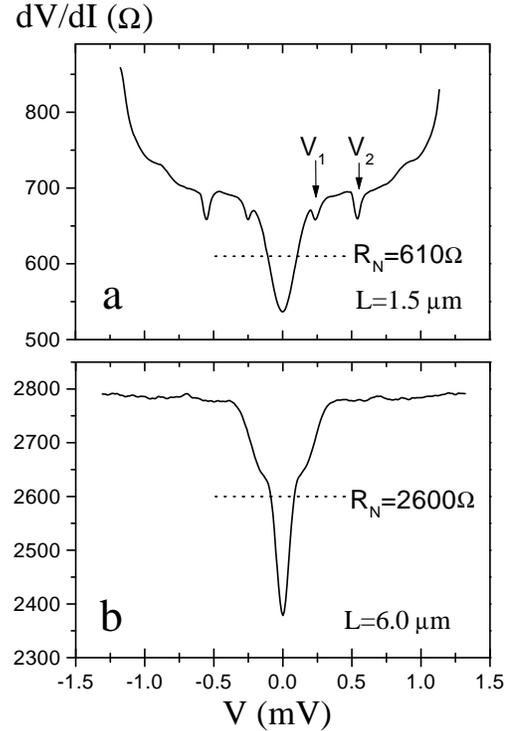}\bigskip}
\caption{Differential resistance versus dc bias voltage for two
samples with the same width $W=0.3~{\mu}$m and different lengths 
(a)~$L=1.5~{\mu}$m and (b)~$L=6.0~{\mu}$m, measured at $T=35$~mK.}
\label{fig2}
\end{figure}

\begin{figure}
\centerline{\epsfxsize2.5in\epsfbox{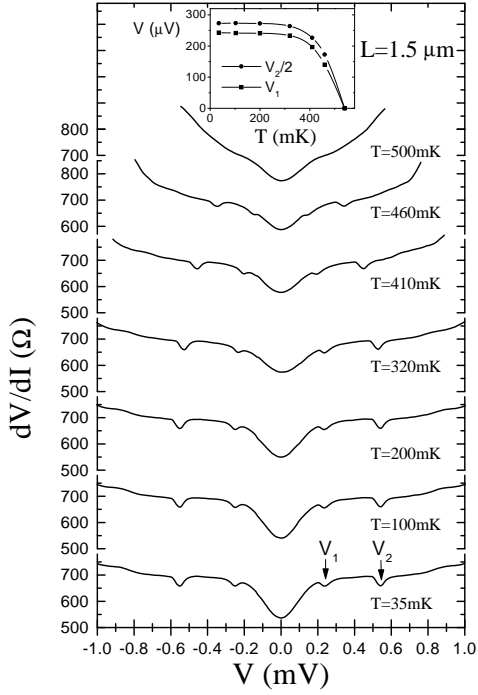}\bigskip}
\caption{Differential resistance of the sample 
with $1.5~{\mu}$m wire length
as a function of bias voltage at different temperatures showing a
zero bias resistance dip and symmetrical dips marked as 
$V_1$ and $V_2$
on the lower curve. Inset: Voltage positions of the resistance dips as a
function of temperature.}
\label{fig3}
\end{figure}

The measurements were carried out with the use of a phase sensitive
detection technique at a frequency of 10~Hz
that allowed us to measure the differential
resistance ($R_{\mathrm{SNS}}=dV/dI$) as a function of the dc voltage ($V$).
The ac current was equal to 10~nA.
Figure~\ref{fig2}
shows typical dependences of $dV/dI$-$V$ for the structures with (a)~short
($L=1.5~{\mu}$m) and (b)~long ($L=6.0~{\mu}$m) wires at~$T=35$~mK. 
Manifest zero bias resistance dip
with respect to the resistance $R_N$ at $T>T_c$ are observed,
with quick initial increasing differential
resistance being followed by a less steep increase at some dc bias.
For SNS junctions 
with the short wires ($L=1.5~\mu$m)
a number of symmetrical features (marked as $V_1$ and $V_2$ in
Fig.~\ref{fig2}a) can be seen. 
It is necessary to note that the resistance value in this bias
range exceeds the wire resistance $R_N$ at $T>T_c$. 
The dependences of $dV/dI$
versus $V$ at different temperatures are shown in Fig.~\ref{fig3}. 
Both the deep minimum at zero bias and the sharp minima at
nonzero biases are temperature dependent. Above $T_c$ the features
disappear. The symmetrical sharp dips at nonzero biases observed 
in $R_{\mathrm{SNS}}$ versus $V$ (Fig.~\ref{fig2}a and 
Fig.~\ref{fig3}) are the so-called
subharmonic energy gap structure (SGS) originated from the multiple
Andreev reflections~\cite{9,10}. The positions of these dips are determined
by the condition $V=\pm2\Delta/en$, with ($n=1,2,3,\ldots$), where $\Delta$ 
is the superconducting energy gap. The dips corresponding to 
$V=\pm2\Delta/e$ and 
$\pm\Delta/e$ manifest in our case. 
The inset to Fig.~\ref{fig3} shows the
temperature dependence of the positions of these dips. It actually
reflects the dependence of the superconducting gap $\Delta(T)$ and strongly
supports the SGS nature of the dips. 
The results presented above show
that SGS is clearly observed even in the case of diffusive transport
with very small mean free path (we have $L/l\sim1.3\times10^3$ for the short
wire) due to a large phase coherence length. 
This shows that in the SNS junctions under study the SGS
is determined by phase coherent transport of retroreflected holes in
the normal wire between the superconducting seas.
The importance of the phase
coherence to observe SGS is supported by the experiment with the long
wires (Fig.~\ref{fig2}b) where no SGS is seen. 
Similar results have been reported recently by
Kutchinsky et al.~\cite{7} where SGS have been studied in Al-$n^+$GaAs-Al
mesostructures.

\begin{figure}
\centerline{\epsfxsize2.6in\epsfbox{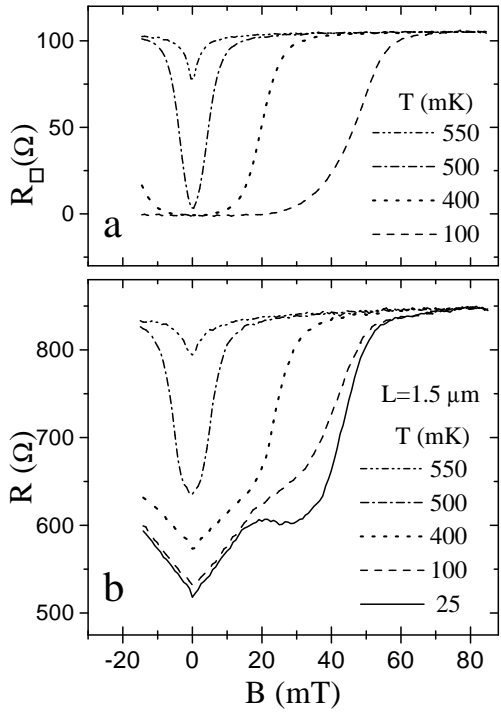}\bigskip}
\caption{(a)~The magnetic field dependences of the sheet resistance
of the 6~nm thick PtSi film measured at several temperatures. (b)~The
zero bias resistance of the sample 
with $1.5~{\mu}$m wire length under the same
conditions.}
\label{fig4}
\end{figure}

Figure \ref{fig4} shows: (a) the superconducting transitions 
of the PtSi film as
functions of the magnetic field at several temperatures, providing the
upper critical fields at various $T$, 
and (b) the zero bias resistance for the structure with short wire
under the same conditions. 
Three distinct regions most pronounced for the curve at the lowest
temperature can be observed in Fig.~\ref{fig4}b: 
(1) at the magnetic fields $B<20$~mT, the SNS junctions
exhibit a linear dependence of $R(B)$, with this feature surviving
up to $T=400$~mK; (2) there is a range of the magnetic fields
where $R=R_N$; (3) a sharp rise of the resistance is resulted from the
transition of the superconducting seas into a normal state. This can
be seen by comparing the results presented in Fig.~\ref{fig4}a and 
Fig.~\ref{fig4}b.

The issue to be addressed now is the behavior of $R_{\mathrm{SNS}}$ 
at the zero magnetic fields and the zero bias. 
As one can see from Fig.~\ref{fig2}, at the
lowest temperatures the value of $\Delta R_{\mathrm{SNS}}/R_N$ for 
all samples is approximately 10\%, with $R_{\mathrm{SNS}}$ 
reaching $R_N$ roughly at the same dc bias.
The similar large value of extraconductance $\Delta G$ has been observed
earlier in mesoscopic SNS and SN junctions~\cite{2,3,4,5,6}. 

There are different theoretical models explaining the
extraconductance, ranging from pair current proximity effect to weak
localization effects. 
These approaches are not necessarily opposed to
each other, but determined by the object under study. 
The analysis of
all these mechanisms showed that our experimental data 
are best explained by the pair current proximity effect. 
If weak localization were the dominant mechanism~\cite{7,11,12,13}, 
we should expect values of $\Delta G$ less than 
$10^{-6}~\Omega^{-1}$ because of very small mean free path.
However, in our experiments we have observed 
$\Delta G\sim2.2\times10^{-4}~\Omega^{-1}$
for the junctions with short wires and 
$\Delta G\sim3.6\times10^{-5}~\Omega^{-1}$ 
for those with long ones. 
The conventional proximity effect is known to imply that
the Cooper pair amplitude decays exponentially with distance into a
normal metal over the characteristic length 
$\xi_N=\sqrt{hD/2\pi kT}$. 
It can effect in decrease the wire length, and resistance.
In our case estimated length $\xi_N\approx 150$~nm at 35~mK,
and we get extraconductance $\Delta G=4\times10^{-4}~\Omega^{-1}$ 
at $L=1.5~\mu$m and $\Delta G=2\times10^{-5}~\Omega^{-1}$  at $L=6.0~\mu$m, 
which is close to $\Delta G$ experimentally
observed at this temperature. We suppose it is the suppression
of the proximity effect that is responsible for reaching $R_N$ on the
dependences of $dV/dI$-$V$ (Fig.~\ref{fig2}) 
and weak plateau on the magnetic
field dependences (Fig.~\ref{fig4}). 
Similar plateaus on the $dV/dI$-$B$ curves were
observed in Ref.~\onlinecite{2}. The following 
increase of the differential resistance
may be connected with the penetration of the normal state into the
superconducting region. 
This phenomenon, that was theoretically
predicted by Geshkenbein et al.~\cite{14}, implies that if 
the transition temperature changes slowly with a distance near
an SN interface at some value of the current 
the penetration of the electrical field into superconductor occurs
This is likely to be our case, as the
reasons leading to the suppression of the superconductivity in the
wires may produce the transitional region with varying order parameter.

In summary, we have observed for the first time the large ZBA and SGS 
in perfect SNS junctions in the regime of diffusive transport. 
To compare, ZBA was observed without
SGS in Ref.~\onlinecite{2,3,4,5,6}.
On the other hand, van Huffelen \textit{et al.}~\cite{8}
saw SGS in their samples, but no ZBA. 
There is the only paper~\cite{7} where
ZBA and SGS occur in the same junction, but this
ZBA is quite different and more complicated 
than the one observed by us. At the lowest temperature a weak ZBA
(1\%--5\%) occurred at low bias ($V<100~\mu$V), it disappeared
together with SGS when the temperature increased, and after that a
new wider ZBA arose.
The difference in the behavior of the junctions can be related to
the quality of NS interfaces.
There is no guarantee that any kind of barrier is absent between the
superconductor and the normal metal or the semiconductor in
the above references. 
The results of our paper strongly support the fact that
ZBA and SGS can be observed in SNS junctions with a perfect NS
interface.

We are grateful to S.~N.~Artemenko, E.~G.~Batyiev and M.~V.~Entin for
helpful discussions. This work has been supported by EU program
PHANTOMS and by grant No. 5-4 of the program ``Physics of quantum and
wave processes'' of the Russian Ministry of Science and Technology.

\begin{table} [btp]
\begin{center}
\caption{The basic parameters of the PtSi films obtained from Hall
measurements at $T>T_c=0.54$~K and physical quantities deduced 
from them (mean free path $l$, elastic scattering time $\tau$ and
diffusion constant $D$).}
\renewcommand{\arraystretch}{1.5}
\begin{tabular}{ccccccc}
$d$ & $R_{\Box}$ & $n$  
& $\tau$ & $k_Fl$ & $D$ & $l$  \\
(nm) & ($\Omega$) & (cm$^{-3}$) 
& (s) &  & (cm$^2$/s) & (nm) \\
\hline
6 &  104 & $7\cdot10^{22}$ &  $8\cdot10^{-16}$ &   
15 &  6 & 1.2
\end{tabular}
\label{table1}
\end{center}
\end{table}


\end{document}